\documentstyle[aps,floats,psfig]{revtex}
\begin{document}
\twocolumn
\draft

\title{Inelasticity in p-nucleus collisions\\
and its application to high energy cosmic-ray cascades}
\author{G. M. Frichter, T. K. Gaisser and T. Stanev}
\address{Bartol Research Institute, University of Delaware\\
Newark, Delaware 19716, USA}
\date{\today}
\maketitle

\begin{abstract}
We employ a simple multiple scattering model to investigate
the inclusive reaction ${\rm p}+{\rm A}\rightarrow{\rm p}+{\rm X}$ for
projectile momenta in the 100-200 GeV/c range. We find that 
data are consistent with a class of interaction models
in which the stopping power of nuclei is rather low.
We discuss extrapolation to ultra-high energy and 
the application to interpretation of cosmic-ray air showers
at energies up to $10^{20}$~eV.
\end{abstract}

\section{Introduction}

To explore the cosmic-ray spectrum beyond about $10^{15}$~eV requires
ground-based experiments with large effective area and long exposure times
to overcome the increasingly low flux implied by the
steeply falling energy spectrum.   Such air shower experiments
cannot observe the primary particle directly but can only
sample the cascade it generates in the atmosphere.  For this reason, 
obtaining results of astrophysical interest, such as the relative fraction
of different types of nuclei or the fraction of gamma-rays, 
requires extensive Monte Carlo simulation to model the cascades
and interpret the measurements.

A recurring problem is that uncertainties in the input to the
calculations introduce corresponding ambiguities in the interpretations
of the experiments.  A major, and to some extent unavoidable, source
of uncertainty is the modelling of the hadronic interactions at
energies well beyond those explored at accelerators.  In the extreme
case, protons with energies near the Greisen-Zatsepin-Kuz'min
(GZK) cutoff \cite{GZK} at $\sim 5\times 10^{19}$~eV correspond to
center-of mass energies more than two orders of magnitude beyond
the highest energy hitherto available at hadron colliders.
There are uncertainties also in 
the region of the ``knee'' of the spectrum around
$5\times 10^{15}$~eV even though this is approximately
equivalent to the center of mass energy of the Fermilab collider.
There are two reasons that significant uncertainties remain even
at this relatively low energy.  One is that interactions in
the atmosphere involve nuclear targets, and in some cases nuclear
projectiles as well.  The other is that it is the forward
fragmentation region of the collision---largely unexplored
by collider detectors---that primarily determines the rate
of energy deposition that generates the core of the atmospheric
cascades.

The most global properties of minimum-bias hadronic interactions
determine the development of air showers.  These include the cross
section and the inelasticity.  The $\bar{p}p$ 
cross section is directly measured
up to $\sqrt{s}\approx 2$~TeV, and its extension to higher
energy can be obtained by extrapolation of fits
based on Regge theory. \cite{Landshoff}
There is less agreement on how to extrapolate inelasticity
and related quantities that determine the rate at which
energy is deposited in the atmosphere via electromagnetic
subshowers.  We focus on inelasticity in this paper.

We are motivated to study this problem now because of intense
experimental activity and ambitious new proposals aimed at
the highest energy cosmic rays \cite{AGASA,HiRes,Auger,EAS1000,OWL}
as well as highly instrumented hybrid arrays aimed at discovering
the sources of cosmic-rays that give rise to the knee feature
in the spectrum, for example \cite{HEGRA,KASCADE,EASTOP,Tibet,SPAM}.
There is a corresponding interest in simulations as illustrated
by the systematic comparison of several codes undertaken by
the group at Karlsruhe. \cite{Karlsruhe}  By  installing
several hadronic event generators into the same cascade code,
they have isolated differences due to the input physics of
the inteaction models from possible technical differences
in how the cascades are followed.  The latter in principle should
not be sources of uncertainty in any case, being determined
by well-known physics such as energy-loss by ionization, pair-production,
bremsstrahlung, etc.

Qualitatively, the {\em inelasticity} of a 
hadronic interaction is the fraction
of the beam energy not carried off by the fragment of the 
incoming particle.  This fraction of the energy is then available
for particle production, including neutral pions which transfer
energy from the hadronic core of the shower into electromagnetic
subshowers.  
Inelasticity is just one moment of one of the inclusive distributions, but
it is arguably the most significant for cascade development
(next to the inelastic cross section itself) because it determines
the rate at which the initial energy of the cascade dissipates.

For $pp$ collisions there is a 
precise experimental definition that involves the inclusive
cross sections 
for production of protons, neutrons and their antiparticles.
Defining 
\begin{equation}
{d\sigma_N\over d^3p}\;=\left[{d\sigma_p\over d^3p}
\,+\,{d\sigma_n\over d^3p}\,-\,{d\sigma_{\bar{p}}\over d^3p}\,-\,
{d\sigma_{\bar{n}}\over d^3p}\right],
\end{equation}
we have
\begin{equation}
\int{d\sigma_N\over d^3p}\,d^3p\;=\;2\sigma_{\rm inel}
\end{equation}
(because there are two nucleons in the initial state) and
\begin{equation}
\label{elas}
\int\,E\times {d\sigma_N\over d^3p}\,d^3p\;=\;\sqrt{s}\times K_{\rm el}.
\end{equation}
Here  $K_{\rm el}$ is the elasticity, and the inelasticity is
defined as 
\begin{equation}
\label{inel}
I\;=\;\left [\; 1\;-\; K_{\rm el}\;\right ].
\label{eqi}
\end{equation}
For pion-initiated interactions, a precise definition
of elasticity requires a model because of the essential
ambiguity between produced and fragment pions.
At high energy the elasticity defined in Eq. \ref{elas}
is approximated by the integral over the leading nucleon
in the lab frame, which is the definition we use
in the remainder of this paper.

For p-nucleus collisions we follow the work of Refs. \cite{Hwa84,Hufner84}
and consider partial inelasticities in the framework of
a Glauber multiple scattering formalism \cite{Glauber70}.
The elasticity is given by
\begin{equation}
K_{\rm el}\;=\;\langle E\rangle\;=\;\sum P_\nu\;\langle E\rangle_\nu,
\label{eqkel}
\end{equation}
where
\begin{equation}
P_\nu\;=\;{\sigma_\nu^{\rm pA}\over \sigma_{\rm inel}^{\rm pA}}
\label{eqpnu}
\end{equation}
is the probability for exactly $\nu$ wounded nucleons in a target
of mass $A$ and $\langle E\rangle_\nu$
is the mean energy of the leading nucleon in collisions with
exactly $\nu$ wounded nucleons.
The partial inelasticity coefficient is defined by the relation
\begin{equation}
\label{partial}
\langle E\rangle_\nu \;=\;(\,1\,-\,I_\nu)\times\langle E\rangle_{\nu-1}.
\end{equation}

The total inelastic hadron-nucleus cross section is
$\sigma_{\rm inel}^{\rm pA} = \sum \sigma_\nu^{\rm pA}$,
and the mean number of wounded nucleons is
\begin{equation}
\langle\nu\rangle\;=\; A \;{{\sigma_{\rm pp}} \over {\sigma_{\rm pA}}}.
\label{eq0}
\end{equation}
We calculate both  $\sigma_{\rm inel}^{\rm pA}$ and   
the partial cross sections $\sigma_\nu^{\rm pA}$
from the cross sections for $pp$ scattering
and the nuclear profiles as described in Ref. \cite{Engel}.

As emphasized in Ref. \cite{Hufner84}, there is no basis
for a naive interpretation of Eq. \ref{partial} because 
fast fragment(s) of the projectile do not reach an asymptotic
physical state until well outside the nucleus.  The strategy
is to assume that $I_1$ is determined by $pp$ scattering and
to treat the remaining partial inelasticities ($\nu > 1$)
as free parameters
constrained by fitting $p$-nucleus data within
the framework of the model.  Our method and conclusions
are similar to those of Ref. \cite{Hufner84}, although
we have improved on their analysis by
using a larger data sample, by treating neutrons and protons
separately and by considering effects of diffraction.

The paper is organized in the following way. Section 2 describes
in detail the multiple scattering model we use for the description of 
 proton-nucleus interactions. Section 3 introduces the fits to $pp$ data
 that we need for the definition of $I_1$ and other parameters
 for the case $\nu=1$.  In section 4 we give
 the fits to p-nucleus data and the resulting
 values for $I_{\nu > 1}$.
 Section 5 contains a discussion of the results 
 in the context of models in current use for calculations
 of cosmic ray cascades at extremely high energies.
 and examples of
 estimated inelasticities in proton air collisions at very high energy.
 Section 6 gives a summary of the results and conclusions.

\section{The Multiple Scattering Model}

The outgoing nucleon in the reaction
${\rm p}+{\rm A}\rightarrow{\rm N}+{\rm X}$ (N being either neutron or proton)
can be specified by its transverse and longitudinal momenta, $p_{\rm t}$ and
$x=p_{\rm l}/p_o$ where $p_o$ is the incident proton momentum.
We model the differential cross section for this process as a sum over
final state distributions corresponding to definite numbers of wounded
nucleons, $\nu$,

  \begin{equation}
  \frac{d^3\sigma^{{\rm p}+{\rm A}\rightarrow{\rm N}+{\rm X}}}
       {dp_{\rm t}^2 dx}\;=\;
  \sum_{\nu=1}^{\rm A} \;\sigma_\nu^{\rm pA}\; M_\nu^{{\rm N}}(x)\;
  \frac{{b^{\rm N}_\nu(x)}^2}{2\pi}\;
  e^{-b^{\rm N}_\nu(x)p_{\rm t}} \:.
  \label{eq1}
  \end{equation}
The transverse momentum distributions are assumed to be described
sufficiently by an exponential form for fixed values of $x$ and are
specified by the slope functions $b^{\rm N}_\nu(x)$. Longitudinal momentum
 distributions for final state nucleons are contained in the functions
$M_\nu^{{\rm N}}(x)$ which are normalized as

  \begin{equation}
  \int_0^1 dx\; M_\nu^{\rm p}(x)\; = \;n^{\rm p}_\nu \:,
  \label{eq2}
  \end{equation}
and

  \begin{equation}
  \int_0^1 dx \;M_\nu^{\rm n}(x)\; = \;n^{\rm n}_\nu \:,
  \label{eq3}
  \end{equation}
with

  \begin{equation}
  n^{\rm p}_\nu\; +\; n^{\rm n}_\nu\; =\; 1 \:.
  \label{eq4}
  \end{equation}
The numbers $n^{\rm N}_\nu$ express the outgoing nucleon multiplicities
for each number of wounded target nucleons. So $M_\nu^{{\rm p}}$ and
$M_\nu^{{\rm n}}$ give the $x$-distributions and relative numbers
of protons and neutrons
after $\nu$ collisions. Eq. \ref{eq4} expresses the fact that our analysis
follows only the forward outgoing nucleon.
When Eq. \ref{eq1} is integrated
over all final nucleon momenta one recovers the inelastic pA cross
section times the mean nucleon multiplicity as expected.

Experimental data on the processes ${\rm p}+{\rm p}\rightarrow{\rm p}+{\rm X}$
and ${\rm p}+{\rm p}\rightarrow{\rm n}+{\rm X}$ may be used to fix
the $(x,p_{\rm t})$ distributions and the nucleon multiplicities for $\nu=1$.
For larger numbers of wounded nucleons we employ the iterative scheme 
discussed above (Eq. 7), which is similar
in spirit to that used by Hwa \cite{Hwa84} and also 
Hufner and Klar. \cite{Hufner84} 
The longitudinal distributions are related by

  \begin{eqnarray}
  M_\nu^{{\rm p}}(x)&=&\int_x^1 \frac{dy}{y}\;
  [\; S_{\nu-1}^+(y)\;\beta_{\nu-1}\; M_{\nu-1}^{{\rm p}}(x/y)\;\nonumber\\ 
  &&+\; S_{\nu-1}^-(y)\;(1-\beta_{\nu-1})\; M_{\nu-1}^{{\rm n}}(x/y)\; ] \:
  \label{eq5}
  \end{eqnarray}
and

  \begin{eqnarray}
  M_\nu^{{\rm n}}(x)&=&\int_x^1 \frac{dy}{y}\;
  [\; S_{\nu-1}^+(y)\;\beta_{\nu-1}\; M_{\nu-1}^{{\rm n}}(x/y)\;\nonumber\\
  &&+\; S_{\nu-1}^-(y)\;(1-\beta_{\nu-1})\; M_{\nu-1}^{{\rm p}}(x/y)\;] \:.
  \label{eq6}
  \end{eqnarray}
The superscripts $+$ and $-$ above describe interactions which
preserve and change the projectile isospin respectively, with the parameters
$\beta$ specifying the fraction of isospin preserving reactions. After $\nu-1$
collisions, a nucleon having longitudinal momentum fraction $x/y$ has
probability $S_{\nu-1}^{+,-}(y)$ to transition to a state having momentum
fraction $x$.

For the case of an incident proton beam, we can take the nucleon 
distributions after zero collisions as boundary conditions,

  \begin{equation}
  M_0^{{\rm p}}(x)\; =\; \delta(1-x) \:
  \label{eq7}
  \end{equation}
and

  \begin{equation}
  M_0^{{\rm n}}(x)\; =\; 0 \:
  \label{eq8}
  \end{equation}
and require that Eqs. \ref{eq5} and \ref{eq6} reproduce the experimentally
determined distributions $M_1^{{\rm p}}(x)$ and $M_1^{{\rm n}}(x)$.
One immediately finds

  \begin{equation}
   S_0^+(y)\;=\;\frac{M_1^{{\rm p}}(y)}{\int_0^1 dy\; M_1^{{\rm p}}(y)} \:,
  \label{eq9}
  \end{equation}

  \begin{equation}
   S_0^-(y)\;=\;\frac{M_1^{{\rm n}}(y)}{\int_0^1 dy\; M_1^{{\rm n}}(y)} \:,
  \label{eq10}
  \end{equation}
and $\beta_0=n^{\rm p}_1$.

Generalizing this result to allow for different inelasticities upon subsequent
collisions, we adopt the power law form with a set of adjustable
parameters, $\alpha_\nu$, to be determined by fits to pA data.

  \begin{equation}
  S_\nu^{+,-}(y)\; =\; \frac{y^{\alpha_\nu}\; M_1^{{\rm p},{\rm n}}(y)}
  {\int_0^1 dy\; y^{\alpha_\nu}\; M_1^{{\rm p},{\rm n}}(y)} \:
  \label{eq11}
  \end{equation}
When $\alpha_\nu=0$ we recover a `naive' multiple scattering model in which
all collisions proceed equally like isolated pp events. With $\alpha_\nu > 0$
contributions from non-leading collisions ($\nu > 1$) are harder than the
initial collision. We will show that the existing data on pA interactions
strongly supports leading baryon spectra that are significantly harder
for $\nu > 1$ than for $\nu=1$.

In order to define partial inelasticity within our formalism we calculate
the mean value of $x$ after $\nu$ collisions,
$<x>_\nu^{\rm p,n}=\int_0^1 dx x M_\nu^{\rm p,n}(x)$ and relate it to
$<x>_{\nu-1}^{\rm p,n}$ with the ratio giving the elasticity
coefficient for the $\nu^{\rm th}$ collision
(or one minus the {\it in}elasticity coefficient).
Integrating Eqs. \ref{eq5} and \ref{eq6} in this manner one arrives
at the relation,

  \begin{eqnarray}
  &&[\;n^{\rm p}_\nu\; <x>_\nu^{\rm p}\;+\;
   n^{\rm n}_\nu\; <x>_\nu^{\rm n}\;]\; =\;(1 - I_\nu)\nonumber\\
  &&\times \;[\; n^{\rm p}_{\nu-1}\; <x>_{\nu-1}^{\rm p}\;+\;
    n^{\rm n}_{\nu-1}\; <x>_{\nu-1}^{\rm n}\;] \:
  \label{eq12}
  \end{eqnarray}
where the mean inelasticity is

  \begin{eqnarray}
  I_\nu\; &=&\;1\; -\; \beta_{\nu-1} \int_0^1 dy\; y S_{\nu-1}^+(y)\nonumber\\
  &&\;-\;(1 - \beta_{\nu-1}) \int_0^1 dy\; y S_{\nu-1}^-(y) \:.
  \label{eq13}
  \end{eqnarray}

\section{Fits to p + p data}

In the current model, the forms determined for
$M_1^{\rm p,n}(x)$ and $b_1^{\rm p,n}(x)$ by fitting
the available ${\rm p}+{\rm p}\rightarrow{\rm p}+{\rm X}$
and ${\rm p}+{\rm p}\rightarrow{\rm n}+{\rm X}$ data can be thought of
as a set of initial conditions that play a crucial role in what we will
eventually infer about leading baryon inelasticity from the p-nucleus data.
This is true primarily because the $\nu=1$ term of Eq. \ref{eq1}
represents 20-30\% of the cross section even for the heaviest nuclei,
and secondarily due to the connection between $M_{\nu > 1}^{\rm N}(x)$ and
$M_1^{\rm N}(x)$ prescribed by Eqs. \ref{eq5} and \ref{eq6}.
Our fitting procedure is straightforward;
we use Eq. \ref{eq1} with $\nu=1$ and
expand $M_1^{{\rm p},{\rm n}}(x)$, and $b_1^{\rm p,n}(x)$ each in a
finite Taylor series. The coefficients are then adjusted to minimize the
$\chi^2$ per degree of freedom when compared to data.

  \begin{figure}
  \centering
  \mbox{\psfig{figure=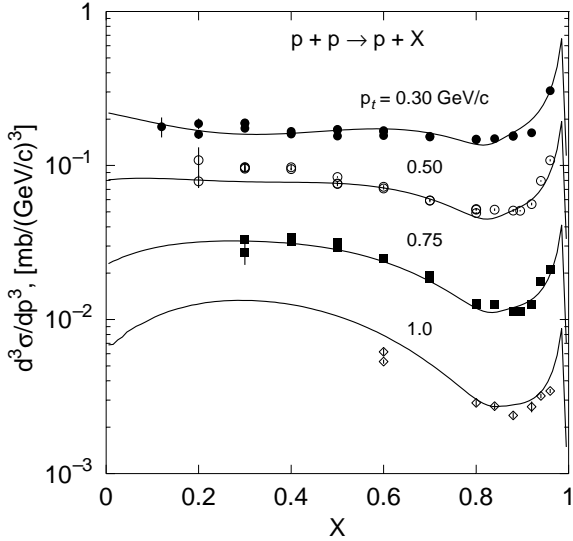,width=8.5cm}}
  \caption{Differential cross section for
   ${\rm p}+{\rm p}\rightarrow{\rm p}+{\rm X}$
   plotted versus momentum fraction $x$ for several transverse momentum bins.
   The data are for beam momenta of 100 and 175 GeV/c
   from references 19) and 20). The curves show our best fit to the data.}
  \label{fig1}
  \end{figure}

Figure (\ref{fig1}) shows the differential cross section for the process
${\rm p}+{\rm p}\rightarrow{\rm p}+{\rm X}$ as a function of longitudinal
momentum fraction, $x$, for several values of the transverse momentum
between 0.3 and 1.0 GeV/c.
The data are for beam momenta of 100 and 175 GeV/c taken
from references \cite{Brenner82} and \cite{Barton83} and
the curves represent our best fit. 

  \begin{figure}
  \vspace*{10pt}
  \centering
  \mbox{\psfig{figure=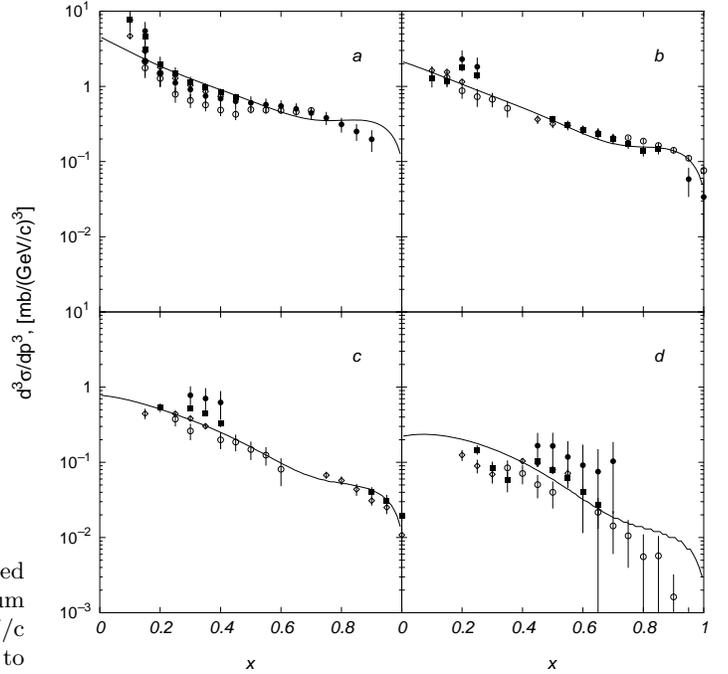,width=13.5cm}}
  \vspace*{0.1truein}
  \caption{Differential cross section for
   ${\rm p}+{\rm p}\rightarrow{\rm n}+{\rm X}$
   plotted versus momentum fraction $x$ for 
   four transverse momentum values: a) 0.15 GeV/c; b) 0.30 GeV/c;
   c) 0.50 GeV/c; and d) 0.75 GeV/c.
   These ISR data correspond to laboratory system beam momenta of 282
   (filled circles), 500 (open circles), 1060 (filled squares), and 
   1500 GeV/c (open diamonds) from Ref. 21.  We have used
   fits of the form of Eq. 9 ($\nu$=1) to interpolate the data
   to selected values of $x$ and $p_T$.  
   The curves show our best fit to the data.}
  \label{fig2}
  \end{figure}

In Fig.~(\ref{fig2}) we plot the differential
cross section for ${\rm p}+{\rm p}\rightarrow{\rm n}+{\rm X}$ 
as a function of $x$ for four
transverse momentum bins from 0.15 to 0.75 GeV/c. The neutron data
correspond to lab system beam momenta of  282, 500, 1060, and 1500  GeV/c
from reference
\cite{Engler75}, and the curves show our best fit. 
Our fits suggest $b_1^{\rm n}(x) \approx b_1^{\rm p}(x)$, so we will not
distinguish between them in the remainder of our discussion.

In fitting the distributions of protons
we have separated the single diffractive and 
non-single diffractive components, so that
$M_\nu^{{\rm p}}(x)=M_\nu^{{\rm p,sd}}(x)+M_\nu^{{\rm p,nsd}}(x)$.
The forward diffractive component (target dissociation) 
represents approximately
10\% of the inelastic $pp$ cross section.  For its $x$-dependence
we use the functional 
form $(1-x)^{-1}$ \cite{Goul83}, with a kinematical cutoff 
near $x=1$.  With
forward diffraction fixed, we then fit $M_\nu^{{\rm p,nsd}}(x)$
by the $\chi^2$ procedure described above. 
In our analysis of $pA$ data below, we make the approximation
of including the diffractive component only in the case $\nu=1$.
This means that diffractive contributions
to pA spectra are confined in our approach primarily to the region 
$x\geq 0.85$.

  \begin{figure}
  \vspace*{10pt}
  \centering
  \mbox{\psfig{figure=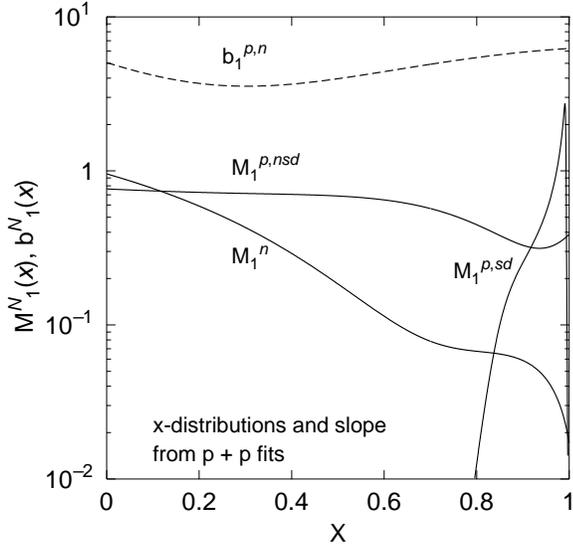,width=8.5cm}}
  \vspace*{0.1truein}
  \caption{Probability distributions for final state protons,
   $M_1^{\rm p,nsd}(x)$, $M_1^{\rm p,sd}(x)$,
   and neutrons, $M_1^{\rm n}(x)$, extracted from
   fits to p + p data. Also shown is the slope function, $b_1^{\rm N}(x)$. The
   mean momentum fractions and multiplicities are quoted in the text.}
  \label{fig3}
  \end{figure}

The functions $M_1^{\rm p,nsd}$, $M_1^{\rm n}$, and $b_1$ obtained from
the best fits are shown in Fig.~(\ref{fig3}) along with the diffractive
component $M_1^{\rm p,sd}$ to complete the picture.
We find for non-single diffractive protons
$<x>_1^{\rm p,nsd}=0.44$ and for neutrons $<x>_1^{\rm n}=0.26$. The proton
and neutron multiplicities derived from our fits 
(including single diffraction) are $n^{\rm p}_1=0.62$
and $n^{\rm n}_1=0.27$. This should be compared with
a proton/neutron ratio of 2:1 in a naive constituent quark
picture of non-diffractive collisions.
Given the overall normalization uncertainties in the pp
data (estimated to be $\sim$20\%), the fitted results are remarkably close.
For the calculation of nuclear processes the distributions
are normalized so that there is exactly one nucleon (proton or neutron)
propagating through the nucleus.  (For the first wounded nucleon
only the normalized distribution includes the diffractive component.)

We note from our fits that final state neutrons appear to be 
significantly softer
than their non-diffractive proton counterparts, although the uncertainties
in the data for $p\,p\,\rightarrow n\,+$~X are larger than
for production of protons.   This difference
in momentum distribution for neutrons and protons 
can have two important consequences.
First, in Eqs. \ref{eq5} and \ref{eq6} there is a `mixing' of
neutron and proton spectra controlled by the parameters
$\beta_{\nu > 0}$. It is easy to see that the amount of this mixing can affect
fits to p-nucleus proton spectra if the input ($\nu=1$) spectra differ. 
We will examine this in the next section.
Second, because the inital neutrons are softer, the leading nucleon 
inelasticity relevant to high energy cascade simulation may be less than
one predicted on the basis of proton data alone together with the
assumption that the inclusive distribution of
neutrons is similar to that of non-diffractive protons.

\section{Fits to p + nucleus data: Inelasticity}

 The number of terms to keep in Eq. \ref{eq1} can be guided
 by considering the Glauber probabilities
 $\sigma_\nu^{\rm pA} / \sigma_{\rm inel}^{\rm pA}$
 for heavy target nuclei (mass $\sim 200$). One finds that roughly
 90\% of the cross section is obtained by the first five terms and
 99\% by the first eleven. We have terminated the sum at $\nu=12$.

 Twelve terms results in a large number of parameters to be fit unless
 some additional assumptions are made. Our approach is to treat all
 interactions subsequent to the initial one on the same footing.
 This means that $\alpha_{\nu-1}$, $\beta_{\nu-1}$, and the functions
 $b_\nu(x)$ have the same value for $\nu > 1$. Indeed we have checked
 that relaxing this constraint has no substantial impact on the results.
 Only marginally better fits are obtained if, for example, we allow a
 different $\alpha$ for each value of $\nu$.

 We give the slope functions the simple quadratic form
 $b_{\nu > 1}(x)=a+bx+cx^2$ and have checked that higher order
 terms do not substantially improve the fits to data. These three
 parameters together with $\alpha_{\nu\geq 1}$ and $\beta_{\nu\geq 1}$
 give a total of five free parameters for fitting the p-nucleus data.

  \begin{figure}
  \vspace*{10pt}
  \centering
  \mbox{\psfig{figure=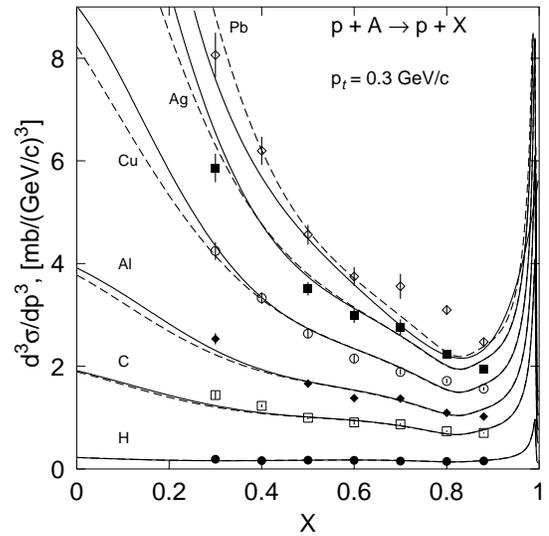,width=8.5cm}}
  \vspace*{0.1truein}
  \caption{Differential cross section for the process
   ${\rm p}+{\rm A}\rightarrow{\rm p}+{\rm X}$
   plotted versus momentum fraction $x$ at fixed $p_{\rm t}=0.3$ GeV/c for
   100 GeV/c protons on targets ranging from hydrogen to lead.
   The data are from reference 20). The dashed and solid curves
   show our best fits with $\beta=2/3$ and $\beta=1$ respectively.}
  \label{fig4}
  \end{figure}

 We have examined two distinct cases for the value of $\beta_{\nu\geq 1}$
 based on different extreme pictures for the nucleon propagation through
 the nucleus. The first can be thought of as the naive case in which all
 interactions proceede identically; that is, the probability for isospin
 preserving reactions at each step is just equal to the proton multiplicity
 observed in p + p reactions, $\beta_{\nu\geq 1} = n_1^{\rm p} = 2/3$.
 Note that in this case the probability that the leading nucleon is
 a proton quickly approaches 1/2 with increasing $\nu$
(14/27 for $\nu=3$ and is ${{3^\nu - 1} \over {2 \times 3^\nu}}$).
 The opposite extreme is that the isospin of the leading nucleon is
 determined solely at the first interaction. This second case corresponds
 to $\beta_{\nu > 1} = 1$. We will show that these two pictures lead
 to somewhat different conclusions for the inelasticity of non-leading
 interactions required to fit the data.

  \begin{figure}
  \vspace*{10pt}
  \centering
  \mbox{\psfig{figure=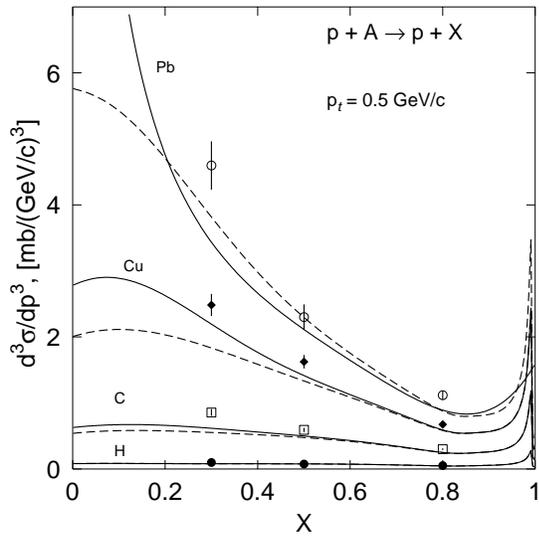,width=8.5cm}}
  \vspace*{0.1truein}
  \caption{Same as Fig. (4) but with $p_{\rm t}=0.5$ GeV/c.}
  \label{fig5}
  \end{figure}

\begin{figure}
  \vspace*{10pt}
  \centering
  \mbox{\psfig{figure=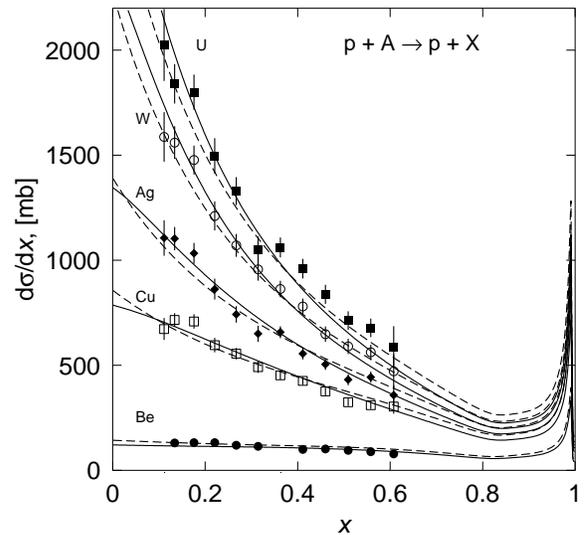,width=8.5cm}}
  \vspace*{0.1truein}
  \caption{Differential cross section $d\sigma / dx$ for the process
   ${\rm p}+{\rm A}\rightarrow{\rm p}+{\rm X}$
   plotted versus momentum fraction $x$ for 120 GeV/c protons on
   targets ranging from Beryllium to Tungsten. The data are from reference
   22). The dashed and solid curves
   show our best fit with $\beta=2/3$ and $\beta=1$ respectively.}
  \label{fig6}
  \end{figure}

 The data we use to study inelasticity are inclusive proton spectra
 from p + nucleus reactions by Barton {\it et al.} \cite{Barton83} and
 Bailey {\it et al.} \cite{Bailey85} with beam energies of 100 and 120 GeV
 respectively. The 100 GeV data were collected for C, Al, Cu, Ag, and Pb
 targets for two transverse momentum bins of 0.3 and 0.5 GeV/c. The 120 GeV
 data were reported summed over transverse momenta for Be, Cu, Ag, W, and U
 targets. These data along with our best fits are shown in Figs.~(\ref{fig4}),
 (\ref{fig5}), and (\ref{fig6}). The dashed and solid curves in these
 figures correspond to the $\beta=2/3$ and $\beta=1$ scenarios respectively.

\begin{figure}
  \vspace*{10pt}
  \centering
  \mbox{\psfig{figure=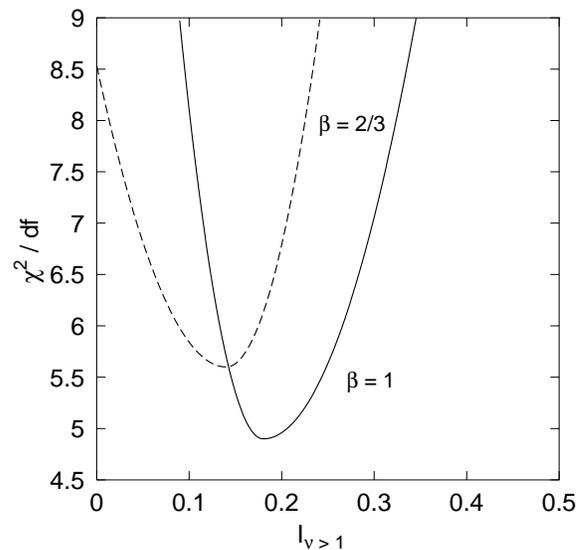,width=8.5cm}}
  \vspace*{0.1truein}
  \caption{Quality if fit, $\chi^2$ per degree of freedom, is plotted versus
   inelasticity of non-leading collisions, $I_{\nu>1}$, 
   for our fits to p + nucleus
   data. Separate curves for the $\beta=2/3$ and $\beta=1$ cases are shown.
   The minima occur at $I_{\nu>1}=14\%$ and $I_{\nu>1}=18\%$, 
   with the latter providing a somewhat better overall fit.}
  \label{fig7}
  \end{figure}

In Fig. (\ref{fig7}) we show the $\chi^2$ per degree of freedom statistic
of these fits plotted as a function of the inelasticity of non-leading
interactions as determined according to Eq. \ref{eq13}.
The $\beta=2/3$ and $\beta=1$ scenarios yield their best fits for $I=$
$.14$ and $.18$ respectively. The $\beta=1$ case offers a somewhat better
overall fit to the data. It is clear that a naive multiple
scattering picture which corresponds to the case of $\beta=2/3$ and $I=.5$ is
certainly not supported in the present analysis.

\section{Implications for cascades}
   
Most hadronic models currently in use for calculation of air showers
at high energy fall into one of two categories.  
One group \cite{Ranft,Werner,Ostapchenko,SIBYLL}
is based on the dual parton model (DPM) \cite{Capella} or the related
Quark-Gluon String model (QGS) \cite{QGS}. 
Another approach is to use some variation of statistical
or thermodynamical ideas \cite{Fermi,Landau}, producing 
particles {\it via} clusters or fireballs, but constrained to agree with
the observed persistence of some high-energy fragments of the projectiles.
There are several examples of this type of model,
for example Refs. \cite{WW,Fowler,Wilk}.  Here we focus
on one particular model \cite{Ding} that has been used recently to
reevaluate implications of the Fly's Eye measurements 
for cosmic-ray composition around $10^{18}$~eV.  The model  
of Ref. \cite{Ding} is an extrapolation to high energy
of the work of Chou and Yang \cite{CNY}.  

In the first group of
models minimum-bias hadronic interactions proceed by the exchange
of strings stretched between fragments of the 
incoming projectile and target particles.  Strings radiate
a characteristic multiplicity of secondaries per interval of
rapidity, so the increase of multiciplicity is essentially
logarithmic in energy (or more accurately, a power of the logarithm
because the number of exchanged strings increases with energy).
Inelasticity is determined basically by the momentum distributions of
the valence constituents, increasing slightly with energy
as more soft strings (coupled to sea quarks rather
than valence quarks) are exchanged.

\begin{table}
\caption{Proton-proton  and
p-air inelastic cross sections with corresponding mean number
of wounded nucleons from Eq. 8.}
\begin{tabular}{|r|rr|r|}
Energy (GeV) & $\sigma_{pp}$ (mb) & $\sigma_{p-{\rm air}}$ (mb) & 
$\langle\nu\rangle$ \\ \hline
 $10^3$    & 33.0 & 284. & 1.69 \\
 $10^7$    & 67.0 & 427. & 2.29 \\
 $10^9$    & 102. & 542. & 2.77 \\
 $10^{11}$ & 142. & 661. & 3.14 \\ \hline
\end{tabular}
\end{table}

\begin{table}
\caption{Inelasticities for proton-proton interactions and
for proton-air interactions for two classes of models:
A=string-type models and B=statistical-type models (see text).
The headings for p-air correspond to different assumed values
of $I_{\nu>1}$, e.g. $I_{\nu>1}=0.14,~0.18$, etc.}
\begin{tabular}{|r|cc|cccc|ccc|}
 & pp &($I_1$) &\multicolumn{4}{c|}{p-air (A)}&\multicolumn{3}{c|}{p-air (B)}\\
Energy (GeV) & A  & B &$I_1$ & .14 & .18 & $I(E)$ & .14 & .18 & $I_1$ \\
\hline
$10^3$    & .50 & .50 &  .55 & .56 & .63 &.53& .55 & .56 & .63 \\
$10^7$    & .55 & .26 &  .62 & .64 & .74 &.58& .38 & .40 & .45 \\
$10^9$    & .57 & .19 &  .66 & .66 & .79 &.60& .35 & .39 & .40 \\
$10^{11}$ & .58 & .15 &  .68 & .70 & .81 &.61& .35 & .39 & .36 \\ 
\hline
\end{tabular}
\end{table}

The cluster models are generally characterized by a more rapid,
power-law dependence of multiplicity on invariant mass of 
the produced clusters.  The observed rise of multiplicity
in the central region is then matched by requiring
the events to become increasingly elastic as energy increases,
so that the fraction of total event energy going into particle
production decreases while the fraction going into
the leading nucleons increases.
(See Ref. \cite{Wlo} for a discussion of inelasticity
in the context of this class of models.)

In Table 1 we show estimates for pp and p-air
cross sections along with the mean number of wounded nucleons per interaction
from Eq. \ref{eq0} for lab energies ranging from ISR to those relevant
in EAS analysis. We note here that the
range of nuclei used in our study of inelasticity, Be ($A\approx 9$)
to Pb ($A\approx 207$) and U ($A\approx 238$), is nicely matched
to the range of energies we wish to consider for hadron
collisions in air.  For a nucleus of mass 200, using a
standard estimate of the p-nucleus cross section, \cite{Denisov}
we find $\langle\nu\rangle_{200}\approx 3.77$ from Eq. \ref{eq0}
at low energy.  In comparison, the mean number of wounded nucleons
expected in a proton-air collision at $10^{20}$~eV from Table 1
is $3.14$.

Extrapolations of the two different types of models
for hadron-hadron interactions beyond collider energies
diverge significantly. We illustrate this in  the first
section of Table 2
by listing the inelasticity for $pp$ collisions, $I_1$, as a 
function of energy to represent the two classes of models.
The $pp$ inelasticity in column A is chosen to be similar
to that of Ref. \cite{SIBYLL}, while that for B is from
the work of Ding et al. \cite{Ding}.  In both cases
we use for illustration the traditional value of 0.5
at low energy, rather than the somewhat higher value
implied by the distributions in Fig. 3.

Next we calculate the corresponding inelasticities for
p-air collisions in the two classes of models starting from
the the assumed values of $I_1$ and 
using Eqs. \ref{eqi}, \ref{eqkel}, \ref{eqpnu} and \ref{partial} 
to calculate overall inelasticity for various assumption
about $I_{\nu>1}$.  Based on the analysis of this paper,
we use $I_{\nu >1}=$ 0.14, 0.18.  
We also show the result of the `naive' model
of propagation through the nucleus ($I_{\nu >1}=I_1$) for illustration,
although we have seen that it is inconsistent with existing data.
(The column labelled I(E) is discussed below.)  
At the highest energies,
we see that `statistical' models predict characteristic energy losses per
collision of only 35 to 40\% compared with about 60 to 70\% for `string' models.
Even at energies characteristic of the ``knee'' region, the models are already
predicting significant differences in energy deposition rates for
proton initiated air showers.

The DPM and QGS models incorporate scattering on nuclear
targets explicitly.  For example, when only one target nucleon
is wounded, a constituent quark(di-quark) belonging to the projectile
proton couples to a string that in turn connects to a di-quark(quark)
belonging to the wounded nucleon.  In cases where there
are two or more wounded nucleons in the target, the additional
nucleons are coupled only to the sea quarks of the projectile.
In this way the desired physics is reproduced by the model.
In particular, the excited nucleon, being off mass-shell, does not
interact repeatedly as a physical nucleon inside the nucleus.
Moreover, the extra multiplicity characteristic of a collision
on a nuclear target is naturally confined to the central region
and the target fragmentation region of phase space.
Capella {\it et al.} \cite{Capella} point out that in
their model the partial inelasticity ($I_{\nu>1}$) is of
order $0.2$, decreasing slightly on successive collisions
in the same nucleus.  We have checked that SIBYLL \cite{SIBYLL} also
shows this behavior.  Thus the string-type models are
consistent with the result of our analysis of proton-proton and
proton-nucleus collisions.  In addition, we note that 
in this type of model it may be more natural to make
the choice $\beta = 1$; that is, to assume that the 
ultimate identity of the final state nucleon is
determined only once during the interaction with the nucleus.

  \begin{figure}
  \vspace*{10pt}
  \centering
  \mbox{\psfig{figure=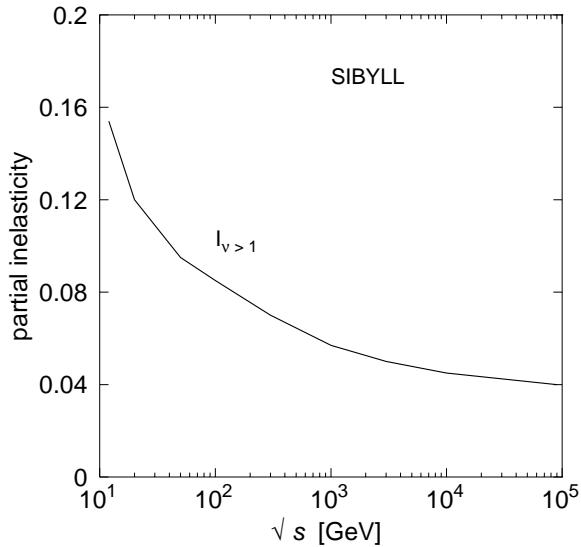,width=8.5cm}}
  \vspace*{0.1truein}
  \caption{Mean partial inelasticity, $I_{\nu > 1}$, versus energy from the
   SIBYLL interaction model. The decrease with increasing energy is 
   intrinsic to string-type models and is a
   consequence of the $x^{-1}$ singularity in the momentum distribution
   of sea quarks.}
  \label{fig8}
  \end{figure}

We note here that the singular nature of the sea quark distributions
for small momentum fractions leads to a threshold effect  
in string-type models for $I_{\nu>1}$.  Asymptotically
the sea-quark momentum on the projectile side becomes
negligible so that the fractional momentum removed
from the projectile by wounded nucleons with $\nu>1$
is small at high energy.  This leads to a
decrease in the value of $I_{\nu > 1}$ as energy increases. 
This behavior is characteristic of string-type interaction models. 
To illustrate, we use the joint probablility
distribution for projectile partons from the SIBYLL interaction 
model \cite{SIBYLL} to evaluate the fraction of energy removed
from the projectile for different numbers of
wounded nucleons. Fig.~(\ref{fig8}) shows the result for $I_{\nu >1}$,
averaged over different values of $\nu>1$.  We see that
energy losses due to interaction with the quark sea
of the projectile decrease significantly at high energy.
The column in Table 2 labeled $I(E)$ uses these energy 
dependent values of $I_{\nu >1}$ for the estimate of the overall
inelasticity. 

The generalization from $pp$ to $p$-nucleus is not
prescribed in the statistical models, at least not
in the form used by Ding {\it et al.} \cite{Ding}.
Given the observed rapid energy deposition in air showers,
users of statistical models generally adopt the ``naive'' treatment
of inelasticity in nuclei ($I_{\nu >1}=I_1$) to compensate for the
intrinsically high degree of elasticity of the hadron-hadron model.
It has also been used in the context of some quark models of
hadron-hadron interactions \cite{KNP}, making the
hadron-nucleus interactions highly inelastic.

\section{Summary}

Analysis of particle production in proton-proton and proton-nucleus 
collisions within a multiple scattering framework leads to the 
conclusion that the second and higher interactions
of the excited nucleon inside the nucleus are relatively
elastic.  Assuming this feature of nuclear interactions
persists to high energy, we can estimate the inelasticity
in hadron-nucleus collisions beyond the energy
range for which we have data.  The results depend on
the behavior of the cross section and inelasticity 
for proton-proton collisions, as illustrated in Tables 1 and 2.
Since the column labelled $I_{\nu>1}=I_1$ is ruled out
by the p-nucleus data, we conclude that the inelasticity
on nuclear targets in the statistical models at high
energy must be quite low.  As pointed out in Ref. 
\cite{Ding}, (see also Ref. \cite{Gaisser}),
such low inelasticity is unable to account well for the
 Fly's Eye data. \cite{Birdetal,Gaisser}  Models of
the type QGS and DPM represent interactions on nuclear
targets in a way that is consistent with the low
energy data on nuclear targets. They predict 
that inelasticity increases slowly with energy,
with a modest increase for nuclear targets.

\acknowledgements
This work was supported in part under Department of Energy Grant Number
DE FG02 91 ER 40626.A007.

\clearpage

\end{document}